\documentclass[12pt]{article}
\usepackage{aaspp4}
\usepackage{epsfig}
 
\begin{document}
\newcommand{\etal}{{et al.~}}

\title{First Structure Formation: \\ 
A Simulation of Small Scale Structure at High Redshift}
\righthead{First Structure Formation} 
\author{Hannah Jang-Condell\altaffilmark{1} and Lars Hernquist\altaffilmark{2}}
\affil{Harvard-Smithsonian Center for Astrophysics}
\authoraddr{60 Garden St., Center for Astrophysics, Cambridge, MA 02138}
\altaffiltext{1}{hjang@cfa.harvard.edu}
\altaffiltext{2}{lhernquist@cfa.harvard.edu}

\begin{abstract}
We describe the results of a simulation 
of collisionless cold dark matter in a $\Lambda$CDM universe 
to examine the properties of objects collapsing 
at high redshift ($z=10$). 
We analyze the halos that form at these early times 
in this simulation and find that the results are similar to 
those of simulations of large scale structure formation at low redshift.  
In particular, we consider halo properties such as the 
mass function, density profile, halo shape, spin parameter, 
and angular momentum alignment with the minor axis.
By understanding the properties of small scale structure formation at 
high redshift, we can better understand the nature of the first structures 
in the universe, such as Population III stars.  
\end{abstract}


\section{Introduction}

The formation of the first stars in the universe, also known as 
Population III stars because of their lack of metals, is important 
for many reasons.  
These objects are responsible for the creation and dispersal of the first 
metals in the universe via Type II supernovae.  The ultraviolet 
radiation from these first stars may also be partly responsible for the 
ionization of the intergalactic medium (\cite{HRL}).  These first objects 
are also the building blocks for the creation of larger structures 
such as galaxies in the bottom-up model of hierarchical structure formation.  

Various workers have studied the process of star formation in the 
early universe using numerical simulations of metal-free gas.  
Spherically symmetric simulations are useful because they are computationally 
less intensive than three dimensional calculations and so can 
include more physics, such as detailed chemistry and cooling 
and even radiative transfer (\cite{HTL}; \cite{omukai+nishi}).  
However, to follow the collapse of matter to stellar densities, 
one needs a fully three dimensional approach.  Several groups have 
carried out such simulations including the relevant chemistry 
(\cite{abel}; \cite{bromm}).

These studies have focused on the evolution of individual density peaks, 
ignoring the effects of other matter in the universe.  
Tidal torques impart angular momentum to collapsing objects, 
affecting their subsequent evolution.  
Virialized halos can merge, changing the structure of objects.  
In this paper, we model a relatively large volume, $1\:h^{-1}$Mpc (comoving), 
that contains a substantial number of halos, in order to understand 
the nature of the first objects that collapse on a more statistical level.  

Since we are interested in the gravitational behavior of the matter, 
we can model the evolution with an N-body simulation.  This is a 
valid approximation for 
a universe where collisionless cold dark matter dominates.
In such a universe, 
the gas is coupled to the dark matter, 
so that gas falls into the potential wells of dark 
matter halos which may in turn lead to star formation.  Thus, 
we can gain some understanding of the sites of the 
formation of the first stars via simulations of dark matter.  

Thus far, numerical simulations with dark matter have focused on the 
problem of large scale structure formation 
(e.g.~\cite{EFWD}; \cite{KHW}; \cite{kauffman}; \cite{laceycole}).
These simulations address the question of structure formation 
on the scale of galaxy clusters in order to understand the processes 
of galaxy formation and clustering.  Here, we apply the 
analytic tools that these groups have developed to address the question 
of structure formation on a smaller scale.

This paper describes the properties of collapsed objects in 
a numerical simulation at high redshift and small scales in 
a $\Lambda$CDM universe.  These objects are some of the first 
non-linear structures in the universe.  The non-linear evolution 
of the power spectrum has been addressed separately (\cite{jang}). 
In \S2, we summarize the computational method used for the simulation;
in \S3, we describe the results of the 
simulation; and in \S4 we summarize our results and 
discuss them in comparison to previous work with N-body simulations.

\section{Method}

In this simulation,
we model the evolution of dark matter in a periodic cube of size
$1\:h^{-1}$Mpc per edge in comoving coordinates.  We adopt a
$\Lambda$CDM cosmology, with $\Omega_m = 0.35$, $\Omega_\Lambda =
0.65$, and $h=0.65$, where the Hubble constant is \(H_0 =
100h\:\)km/s/Mpc.  The power spectrum is normalized to $\sigma_8=0.9$,
where $\sigma_8$ is the rms density variation smoothed with a top-hat
filter of radius $8\:h^{-1}$Mpc.  For the cosmological model we consider,
this normalization is roughly consistent with both the local abundance
of rich clusters (\cite{WEF}) and fluctuations in the cosmic microwave
background as observed by COBE (\cite{bennett}).

Initial conditions were generated using the analytic fit to the CDM 
power spectrum derived by Efstathiou, Bond, \& White (1992)
\begin{equation}
P(k) = |\delta_k|^2 = \frac{Bk}{\{1+[ak+(bk)^{3/2}+(ck)^2]^\nu\}^{2/\nu}}{}\:,
\end{equation}
where
$a = 6.4/\Gamma \:h^{-1}$Mpc, 
$b = 3.0/\Gamma \:h^{-1}$Mpc, 
$c = 1.7/\Gamma \:h^{-1}$Mpc, 
$\nu = 1.13,$
and $B$ is a normalization constant determined by $\sigma_8$.  The
shape parameter $\Gamma$ is set to 
\( \Gamma = \Omega_m h \) for a $\Lambda$CDM cosmology.
This power spectrum 
is extrapolated to a starting redshift of 100
assuming linear theory, and 
is then converted into spatial density fluctuations by first 
assigning random phases to the $\delta_k$ and then taking the Fourier 
transform.  The spatial density fluctuations are converted into 
particle positions and velocities using the Zel'dovich approximation.  

The code used for the simulation is {\tt PTreeSPH}, a gravity treecode with 
smoothed particle hydrodynamics 
(SPH) designed to run on a parallel supercomputer
(Dav\'{e}, Dubinski, \& Hernquist 1997).  The SPH part of the code 
was unused in this simulation since gas was not included.
The code uses a Barnes-Hut (\cite{barneshut}, \cite{hernquist87}) 
algorithm for computational efficiency in calculating 
gravitational forces and a spline kernel for gravitational softening
(\cite{HK}).  The softening length chosen for this 
simulation was $1/20$th of the mean interparticle separation, or 
$1/2560$th of the box size.  
The simulation box contains $128^3$ particles, implying that 
each particle represents $4.6\times10^4 M_{\sun}$ of dark matter.
The simulation was run on a four-processor Beowulf-type cluster of PCs
at the Harvard-Smithsonian Center for Astrophysics.

\section{Results}
\label{results}

\begin{figure}
\epsfig{file=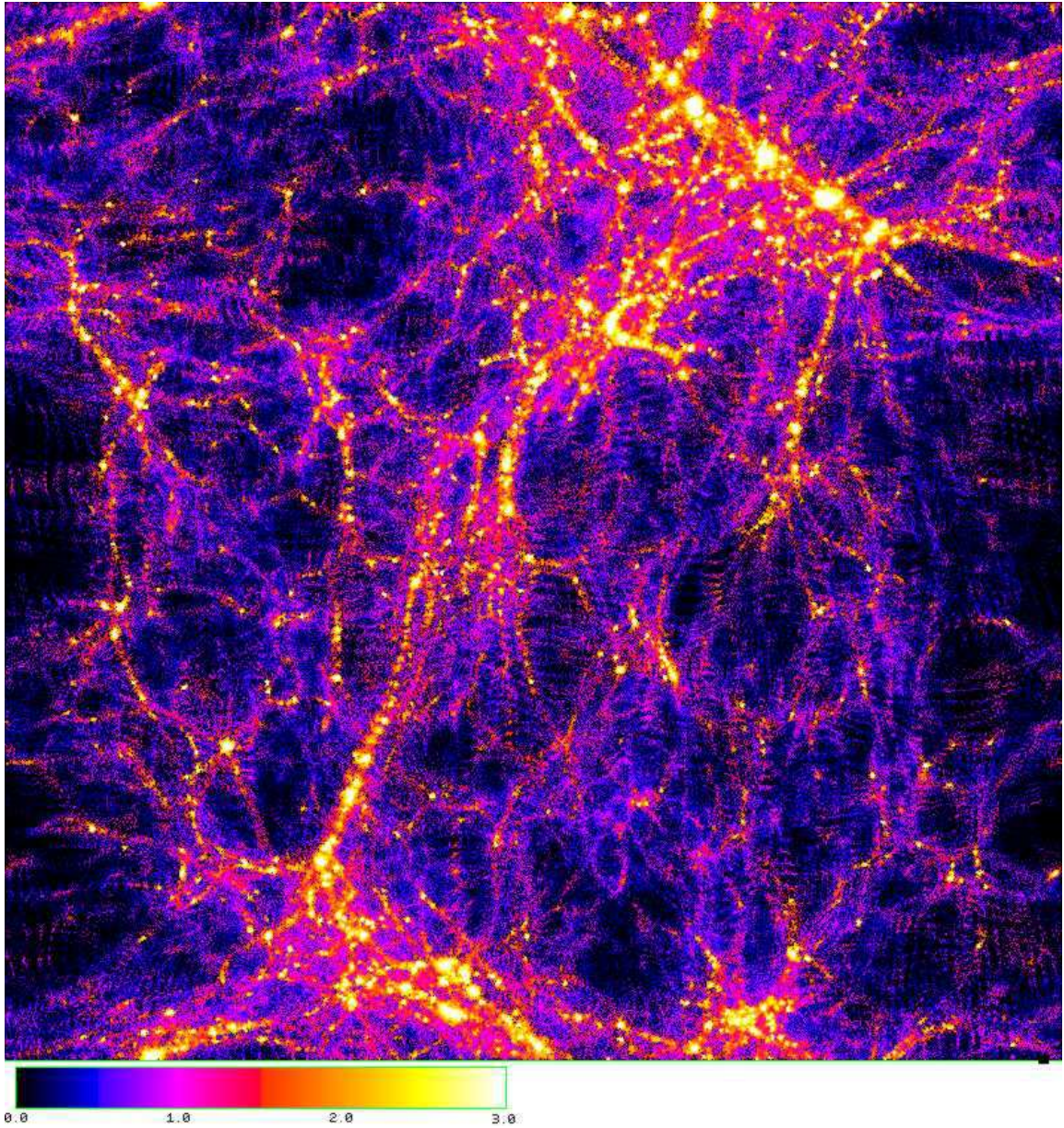,width=6in}
\caption{\label{prettypic}Simulation output at redshift $z=10$.  
The particles are shown in projection along the $x$-axis.  
Color indicates log density in units of $\bar{\rho}$.  The size of 
the box is $1\:h^{-1}$Mpc (comoving) per side.  (This figure is 
also available at 
{\tt http://cfa-www.harvard.edu/\~{}hjang/research/prettypic.gif}
)}
\end{figure}

The particle distribution at the final output of redshift $z=10$ 
is shown in Fig.~\ref{prettypic}.  
The particles are displayed in projection along the $x$-axis and 
the colors indicate the particle densities.  Already at this 
high redshift there is evidence of structure formation in the form of clumps 
and filaments of high particle density.  The structures are similar to 
those seen in large scale simulations evolved to much lower redshift.  

As we can see in Fig.~\ref{prettypic}, 
the centers of dense knots have densities of $\gtrsim1000\bar{\rho}$.  
Density evolves with redshift as $a^{-3}=(1+z)^3$, so at $z=10$,
\( 1000\bar{\rho} = 1000 a^{-3} \Omega_b \rho_{\mathrm{crit}} \)
where $\Omega_b$ is the baryon density of the universe in units of the 
critical density, and $\rho_{\mathrm{crit}}$ is the present day critical 
density.  Constraints on light element production during big bang 
nucleosynthesis indicate that \(0.01 \lesssim \Omega_b h^2 \lesssim 0.015\) 
(e.g.~\cite{peacock}, \S9.5).  Thus,
\begin{equation}
1000\bar{\rho} \approx 3\times10^{-25}\:\mathrm{g\:cm}^{-3}
\end{equation}
which is $\sim 0.2\:\mathrm{cm}^{-3}$ in hydrogen atoms.  This is 
comparable to the density of neutral hydrogen in the present day 
ISM, which is $\sim 1\:\mathrm{cm}^{-3}$ (e.g.~\cite{spitzer}, \S1.1).  
We can see 
that the densest regions in the simulation will be similar in 
density to the present-day ISM, and so are potential sites for 
star formation.  

\subsection{Finding halos}

The particle distribution was analyzed using a program called {\tt skid} to 
determine the positions, masses, and sizes of collapsed halos.  
A description of {\tt skid} can be found at 
{\tt http://www-hpcc.astro.washington.edu/tools/SKID/}.  

The basic algorithm that {\tt skid} uses is as follows:
\begin{enumerate}
\item Calculate densities, and consider only those above a user specified 
density threshold.  These are called the moving particles.
\item Slide the moving particles along the density gradient toward higher 
density.  
\item Continue moving particles until all the particles 
stop moving and are localized in high density regions of a 
user specified size ({\tt eps}).
\item Group together these localized particles using the 
friends-of-friends method with a linking-length of {\tt eps}. 
\item Reject groups with less than a user specified minimum
number of particles. 
\item Particles which are not bound to their group are removed from the group.
\item Reject groups with less than minimum number of particles.
\end{enumerate}

We used a density threshold of $200\bar{\rho}$, 
a linking-length ({\tt eps}) of $1/640$th of the box size or 
1/5 the mean interparticle spacing, and a minimum halo size of eight 
particles.  Using the simulation output at a redshift of $z=10$, 
this resulted in 2881 halos, with the largest halo 
consisting of 8475 particles, corresponding to a mass of 
$4\times10^8 M_{\sun}$ and the smallest with eight particles, 
corresponding to $4\times10^5 M_{\sun}$.
The density cutoff was chosen somewhat arbitrarily, and 
changing its value does not significantly affect the overall 
results presented in this paper.  For example, changing the density 
threshold from $200\bar{\rho}$ to $86\bar{\rho}$ did not alter 
the ensemble properties of the halos.  
Some of the halos identified by {\tt skid}
were subsequently rejected from the sample for 
being unbound, as described in \S\ref{energysection}.  
We have analyzed various properties of these halos and describe our results 
below.

\subsection{Mass function}

The distribution of halos masses can be expressed in terms of the mass
function, $f(M)$, where $f(M)dM$ is the number density of halos of with 
mass between $M$ and $M+dM$.  An analytic prediction for the mass function 
can be obtained using the Press-Schechter formalism (\cite{PS}).  

The Press-Schechter formalism states that the fraction of the universe 
condensed into objects of mass $>M$ is 
\begin{equation}
F(>M) = 1 -\mathrm{erf}\left(\frac{\delta_c}{\sqrt{2}\,\sigma(M)}\right)
\end{equation}
where $\sigma(M)$ is the 
rms density fluctuation smoothed over spheres of mass $M$,
and $\delta_c$ is the critical overdensity.  The critical overdensity 
is defined as follows: an object of 
density $\rho$ in a universe of average density $\bar{\rho}$ 
is collapsed when
\begin{equation}
\delta = \frac{\rho-\bar{\rho}}{\bar{\rho}} > \delta_c.
\end{equation}
We take $\delta_c=1.69$, the canonical value.
The mass function then becomes
\begin{equation}
f(M)\,dM = \left|\frac{dF}{dM}\right|\frac{\bar{\rho}}{M}\,dM
	= \sqrt{\frac{2}{\pi}}\,\frac{\delta_c}{\sigma}\,\bar{\rho}
		\left|\frac{d\;\ln\sigma}{d\;\ln M}\right|
		\exp\left(-\frac{\delta_c^2}{2\sigma^2}\right)
		\,\frac{dM}{M^2}.
\end{equation}

\begin{figure}
\epsfig{file=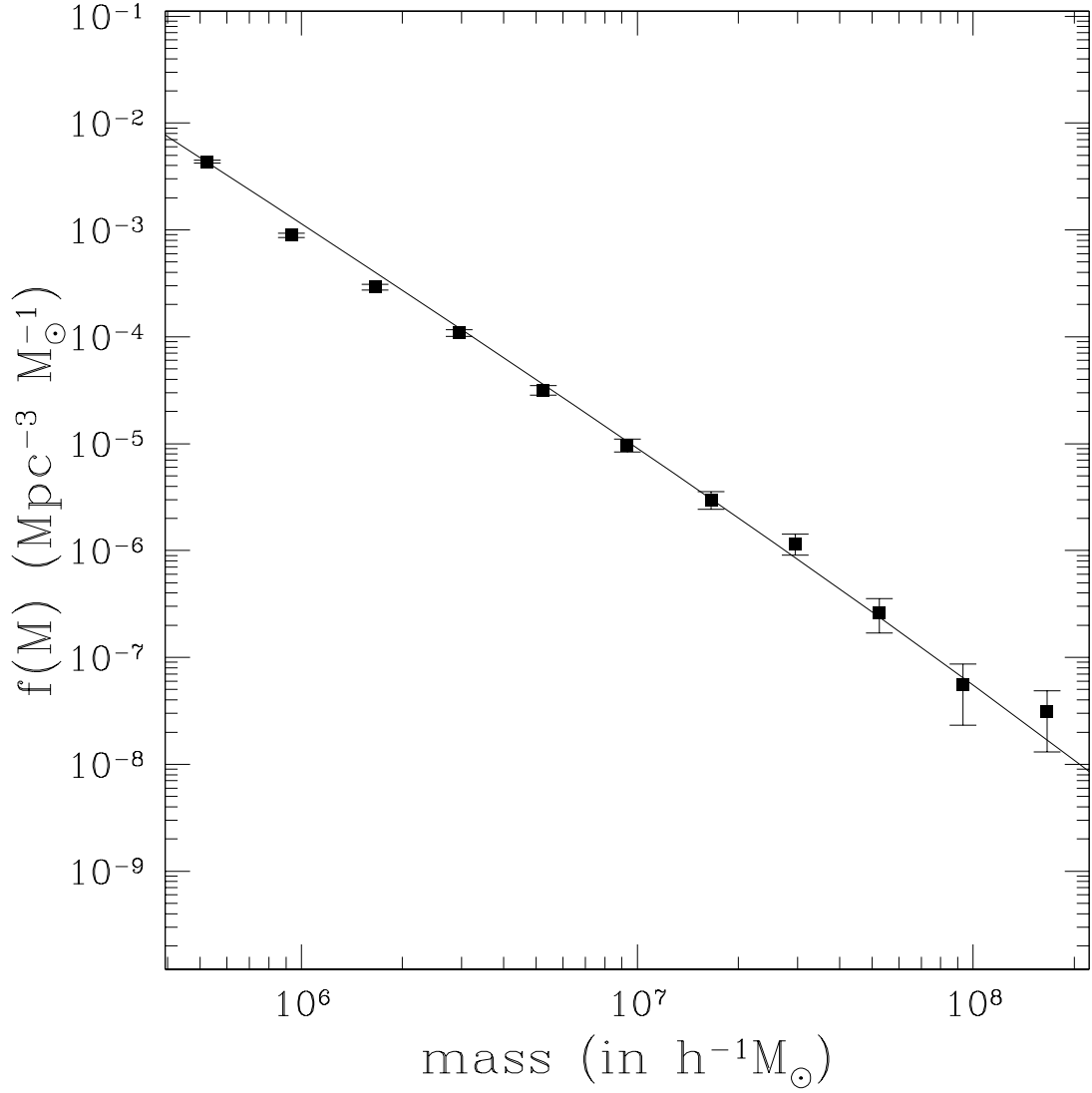,height=6in,width=6in}
\caption{\label{massfunc}Mass function of halos.  The points represent 
the simulation data and the solid line is the Press-Schechter prediction, 
with $\delta_c=1.69$.}
\end{figure}

Figure \ref{massfunc} shows the actual
mass function of halos in the simulation compared to the 
predictions of the Press-Schecter formalism.  Halos with 
positive binding energy were omitted from the calculation 
of the mass function as explained in Section \ref{energysection}.
The results are in remarkably close agreement with the Press-Schechter 
prediction.

\subsection{Halo energies}
\label{energysection}

When {\tt skid} calculates halos from the particle distributions, it 
removes unbound particles from the halos, so in principle, each halo should 
be gravitationally bound.  The binding energy of particle $i$ to the 
rest of the halo is 
\begin{equation}
E_i =  m_i \left[ \frac12 v_i^2 + \sum_{j \neq i} \Phi_{j}(r_{ij}) \right]
\end{equation}
where $m_i$ and $m_j$ are the masses of particle $i$ and particle $j$ 
respectively, $v_i$ is the velocity of particle $i$ with respect to the halo 
center of mass velocity, and $\Phi_{j}(r_{ij})$ is the gravitational 
potential between particles $i$ and $j$.  The gravitational potential 
has the general form \(\Phi_{j}(r_{ij}) = -G m_j f(r_{ij})\), which 
admits a softened potential for particles pairs with small 
$r_{ij}$ (\cite{HK}).

The requirement for binding is that $E_i<0$.  
The binding energy for the halo as a whole is the sum of its gravitational
potential energy and its kinetic energy.  Thus,
\begin{equation}
\label{totalenergy}
E = K+W = \frac12 \sum_i m_i v_i^2 - \frac12 \sum_i \sum_{i\neq j}
	G m_i m_j f(r_{ij}).
\end{equation} 
The factor of $\frac12$ in the potential energy is to account for 
double-counting the interactions between pairs of particles.  
Equation (\ref{totalenergy}) can be rewritten as 
\begin{equation}
E = \sum_i E_i + \frac12 \sum_i \sum_{i\neq j} G m_i m_j f(r_{ij})
	= \sum_i E_i + |W|
\end{equation}
which is {\em greater} than the sum of the individual binding energies of the 
particles.  In other words, it is possible to have a halo where each particle 
is bound to the sum of all the other particles, but the halo as a whole 
is unbound.  So, after {\tt skid} has calculated the halos, the total 
binding energy of each halo is checked, and discarded from the 
sample if it is unbound.  Out of the total of 2881 halos, 380 were 
rejected in this way, leaving 2501 bound halos.  

\begin{figure}
\epsfig{file=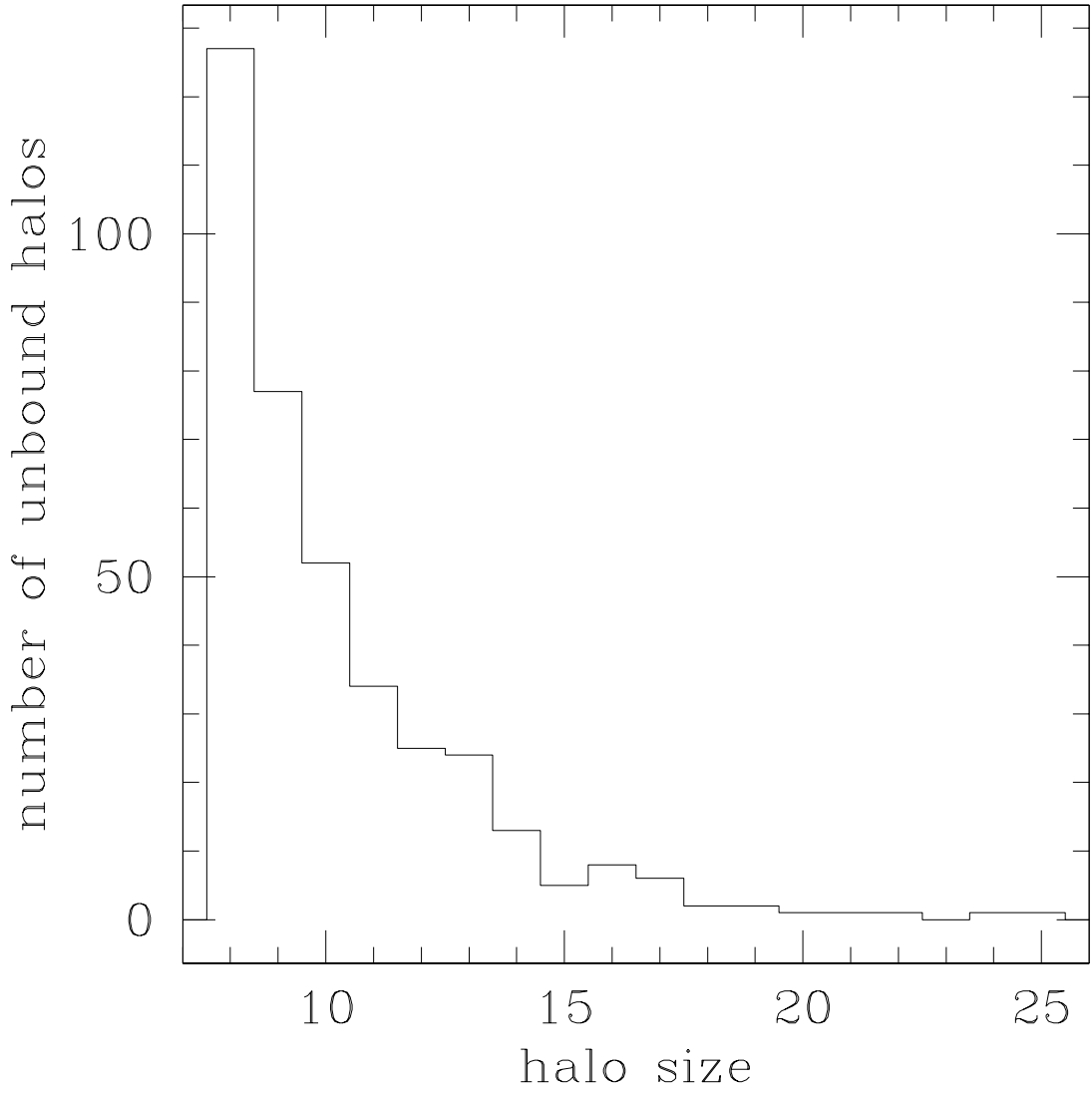,height=6in,width=6in}
\caption{\label{unbound}Distribution of unbound halos with respect to 
halo size in terms of particle number.}
\end{figure}

The distribution in size of the 380 unbound halos is plotted in 
Figure \ref{unbound}.  Note that the unbound halos are all quite small, 
the largest containing only 25 particles.  This shows that our method 
of using {\tt skid} to calculate halos is fairly robust, failing only 
at small halo sizes, which are intrinsically unstable to small 
perturbations.  
In the analysis of the remaining halo properties, calculations are 
done either as a function of mass, or neglecting all halos with less
than 50 particles.  Thus, the removal of these halos 
from the sample will not significantly affect our results.

\begin{figure}
\epsfig{file=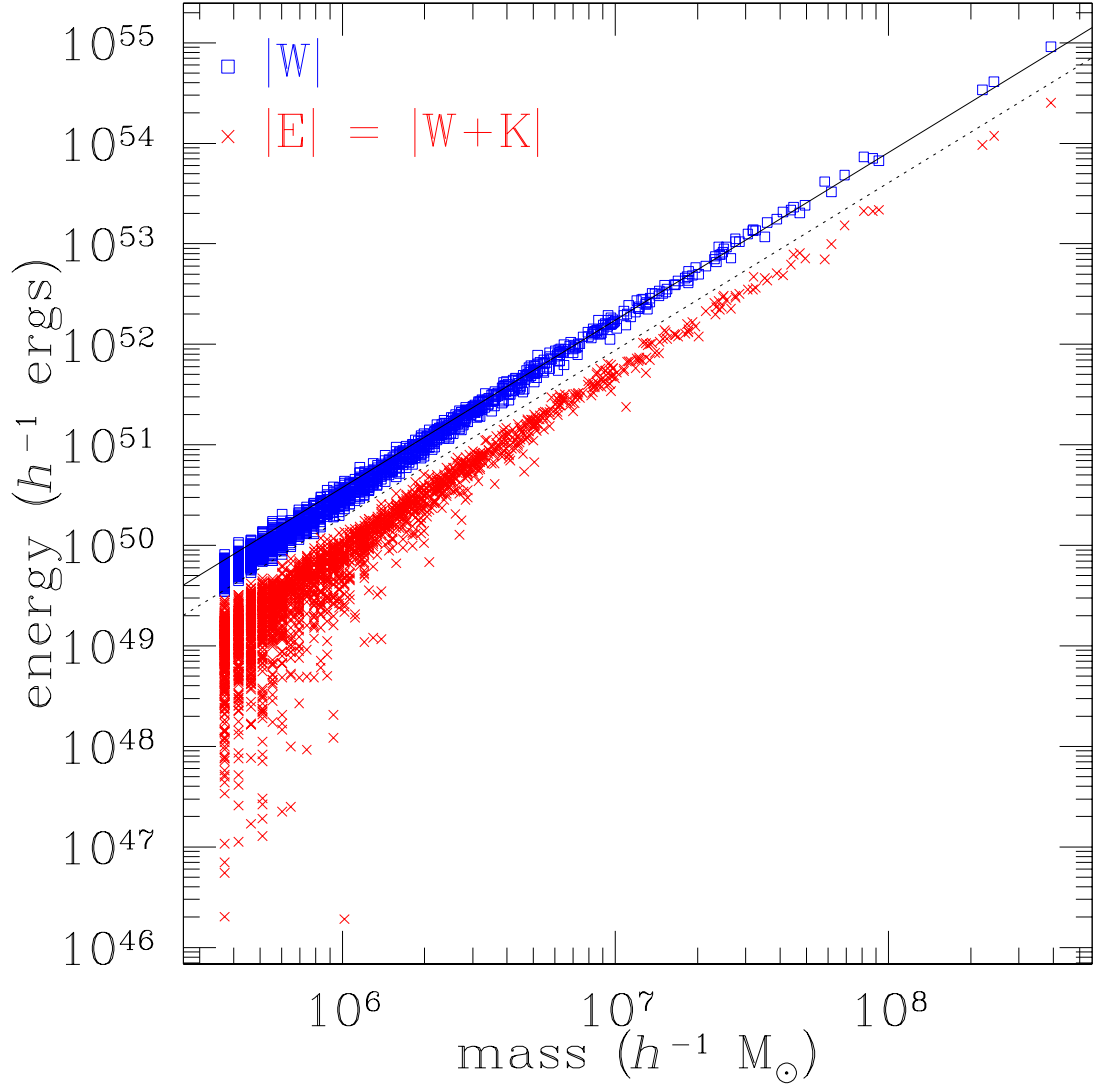,height=6in,width=6in}
\caption{\label{energyplot}Energies of the halos versus mass.  
Potential energy 
is plotted as squares, total binding energy is plotted as crosses.
The solid line is proportional to $M^{5/3}$.}
\end{figure}

Figure \ref{energyplot} is a plot of the energies of the halos versus 
their masses.  The absolute values of the potential energy are plotted as 
squares and the absolute values of the total binding energy are plotted 
as crosses.  The energies all seem to follow a power law of 
\(E \propto M^{5/3}\), as indicated by the solid line.  
This relation can be obtained analytically 
by supposing that the halos are the 
result of the collapse of spheres of uniform 
density.  This is a reasonable assumption since the 
universe was initially very nearly uniform in density.
If the density is constant, then the mass of a sphere of radius $r$ is 
\begin{equation}
M(r) = \frac{4\pi\rho}{3} r^3.
\end{equation}
The equation for the gravitational potential energy
energy of a uniform sphere of density $\rho$ and mass $M$ is 
\begin{eqnarray}
W &=& \int_0^R \Phi(r)\rho\, d^3r \nonumber\\
&=& \int_0^R -\frac{GM(r)}{r}\,\rho\, 4\pi r^2 dr \nonumber\\
&=& -\frac{3GM^2}{5R} \nonumber\\
\label{gravpot}
&=& -\frac35 GM^{5/3} \left(\frac{4\pi\rho}{3}\right)^{1/3}.
\end{eqnarray}
Assuming that the halo is virialized, we have
\begin{equation}
E = K + W = -\frac12 W + W = \frac12 W.
\end{equation}
Thus, the total energy should obey the same scaling. 

\begin{figure}
\epsfig{file=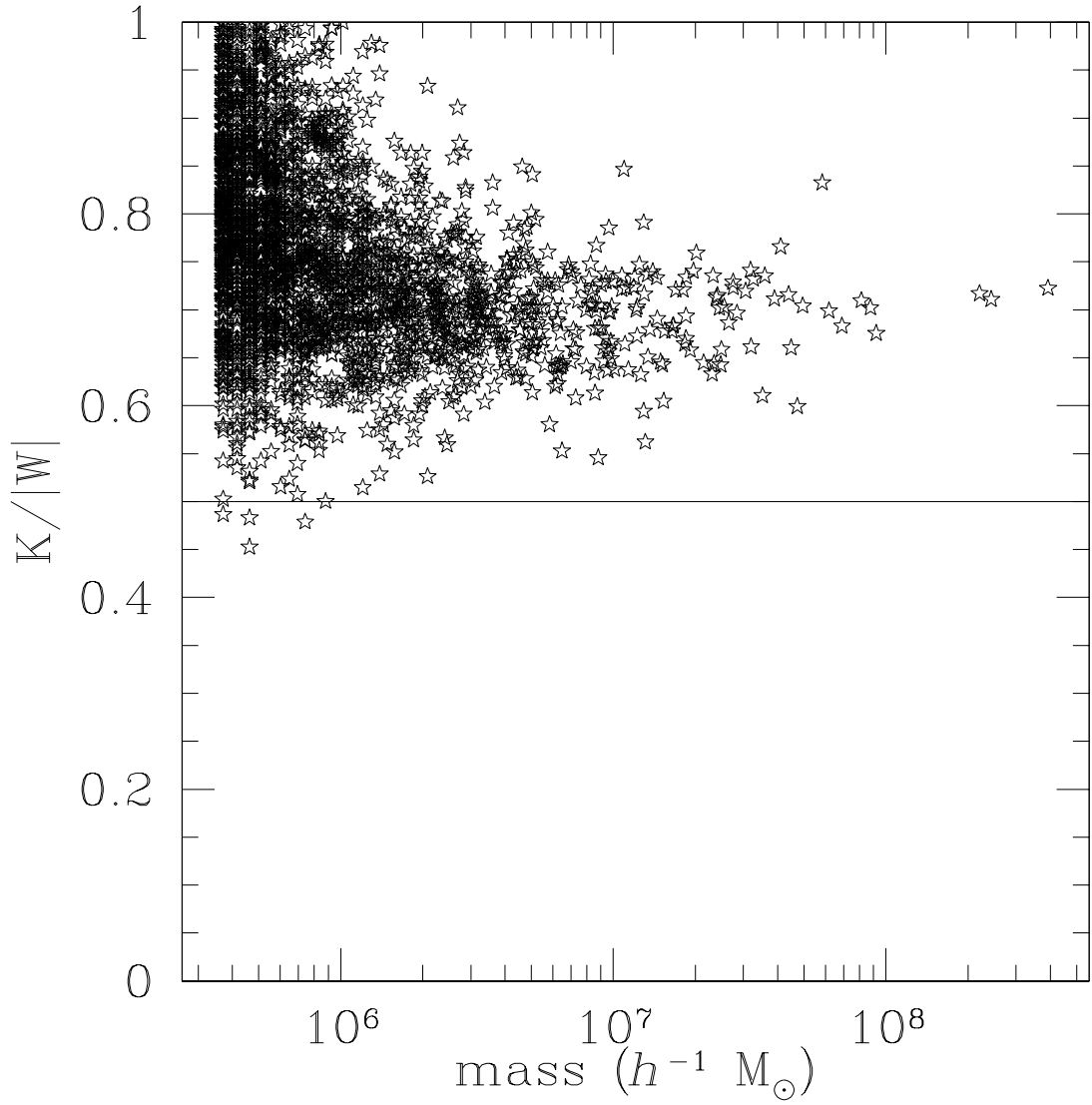,height=6in,width=6in}
\caption{\label{energyratio}Ratio of kinetic to potential energy.  
The solid line indicates $K=\frac12|W|$, which holds for virialized objects. }
\end{figure}

The dotted line represents $1/2$ the energy of the solid line.  
The total energies 
of the halos are all below this line, indicating that $E>\frac12W$ in general.
This means that the halos have not yet become virialized.  
This is also shown in Fig.~\ref{energyratio},
which shows the ratio between the kinetic energy 
and the potential energy.  The solid line marks $K=\frac12|W|$, which 
would indicate a virialized halo.  Nearly all the halos have too high 
a kinetic energy for virialization.

\subsection{Gas temperature}

Assuming that the matter in the universe is dominated by cold dark 
matter, then the gas should trace the distribution of the dark matter. 
In particular, gas should fall into the potential wells of the dark 
matter halos and become shock heated to the virial temperature of
the halo.  We can estimate this temperature 
by assuming that the mean-square velocity of the gas is the same as the 
dark matter.

\begin{figure}
\epsfig{file=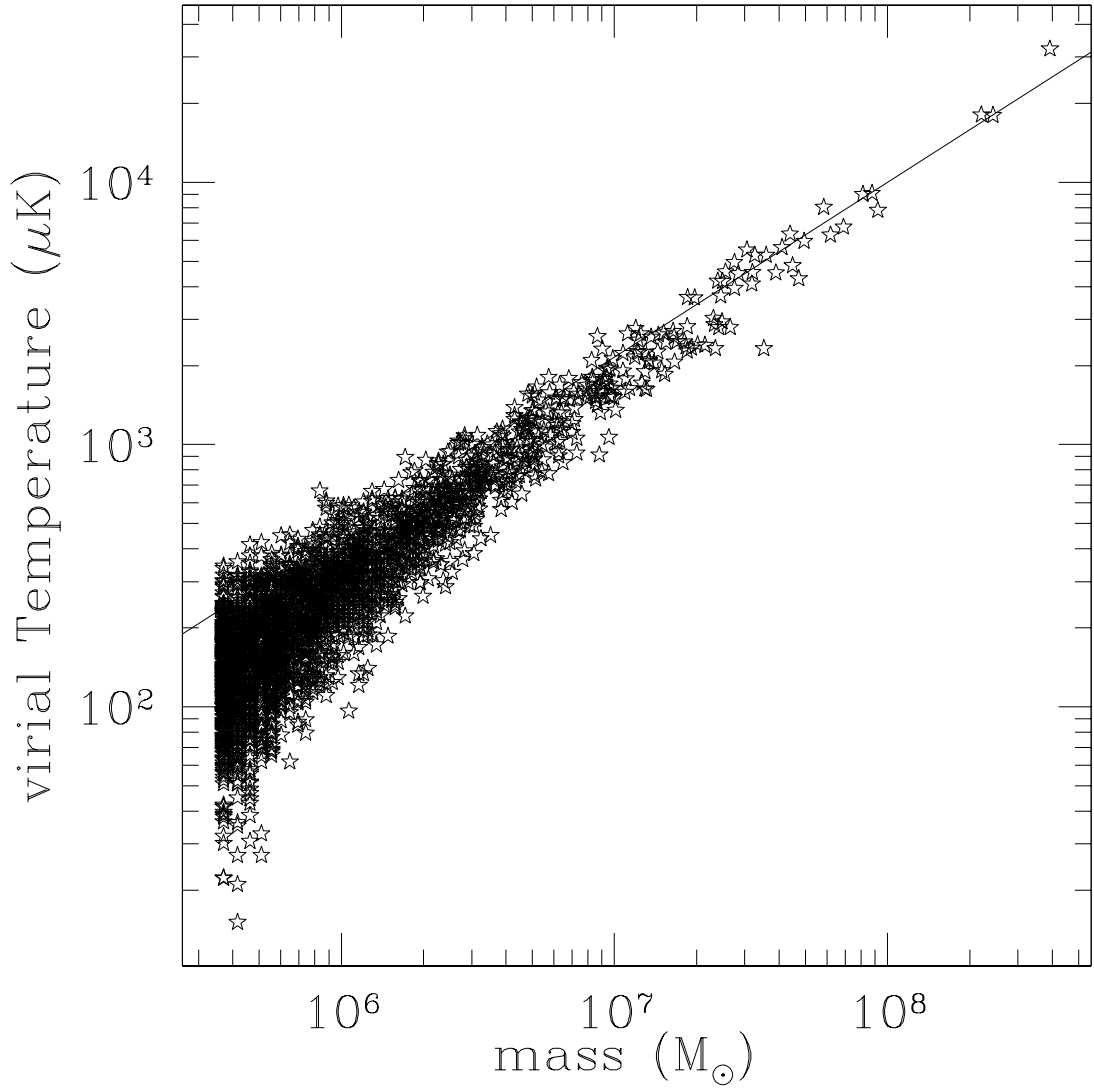,height=6in,width=6in}
\caption{\label{vtemp}Virial temperature versus mass.  The solid 
line is proportional to $M^{2/3}$.}
\end{figure}

For an ideal gas, the temperature is related to the mean-square velocity by
\begin{equation}
\langle v^2 \rangle = \frac{3kT}{\mu m_p},
\end{equation}
where $k$ is Boltzmann's constant, $T$ is the temperature in Kelvin,
$m_p$ is the proton mass, and $\mu$ is the mean molecular weight.  
This temperature, as a function of $\mu$, 
is plotted versus mass for each halo in Fig.~\ref{vtemp}.  
The value of $\mu$ will depend on the composition and ionization 
state of the gas.  

As discussed in section \ref{energysection}, the kinetic energy of the halo 
follows a power law of $K \propto M^{5/3}$.  Since 
\(K = \frac12 M \langle v^2 \rangle \), the temperature should follow
the relation
\begin{equation}
T \sim M^{2/3}.
\end{equation}
This is indicated by the solid line in 
Fig.~\ref{vtemp}.  The temperatures follow this power law relation 
fairly well.

\subsection{Halo profiles}
\label{haloprofile}

\begin{figure}
\epsfig{file=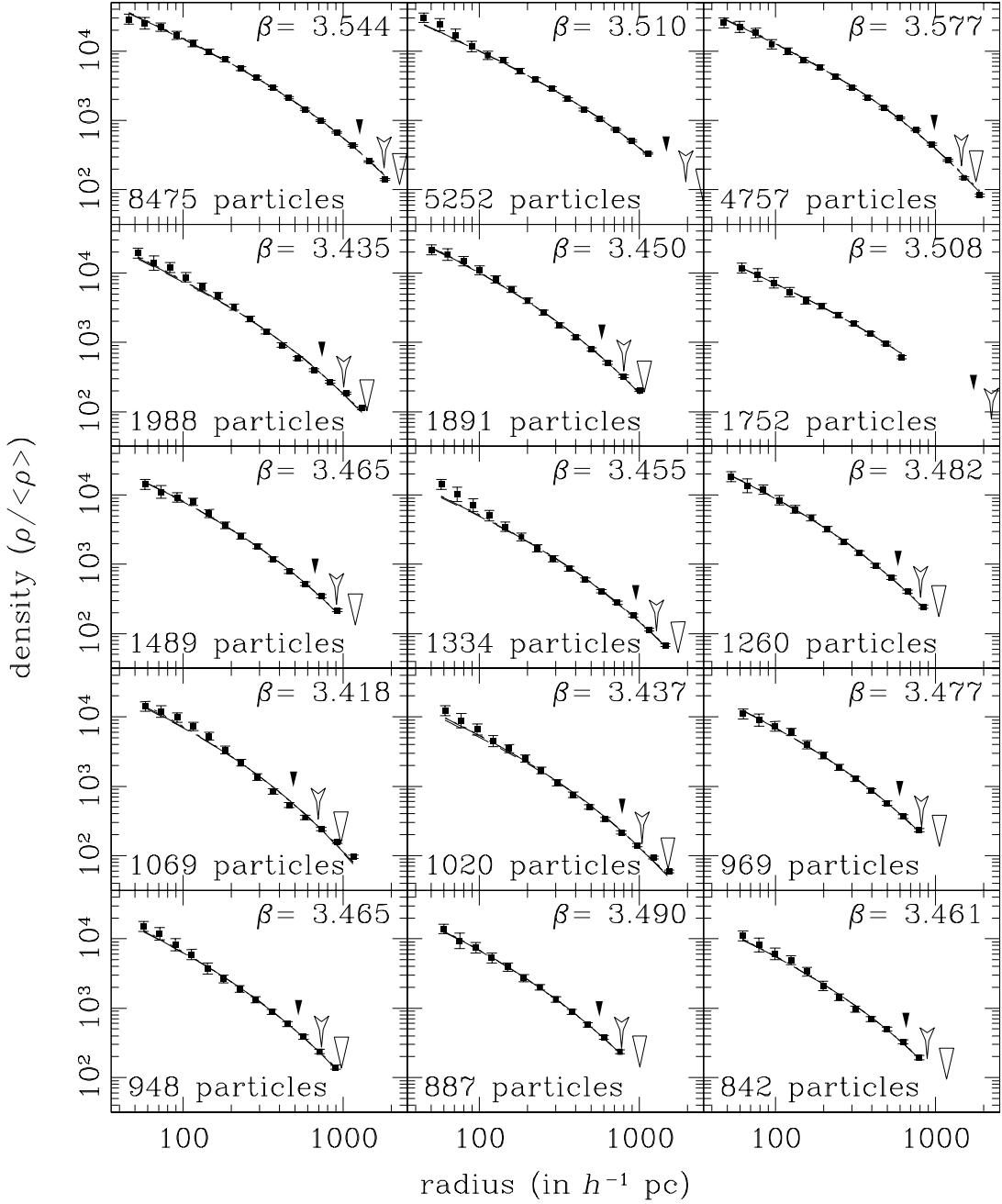,width=6in,bbllx=18,bblly=160,bburx=474,bbury=717}
\caption{\label{profilea}Density profiles of halos.  Radius is 
in physical parsecs.  Dotted line is 
NFW profile fit, short dashed line is Hernquist profile fit, and 
long dashed line is the fit allowing $\beta$ to be a free parameter.
Triangles mark the $r_s$
for each fitted profile -- solid triangle is NFW, hollow triangle 
is Hernquist, and deflated triangle is the best-fit $\beta$ profile.}
\end{figure}

\begin{figure}
\epsfig{file=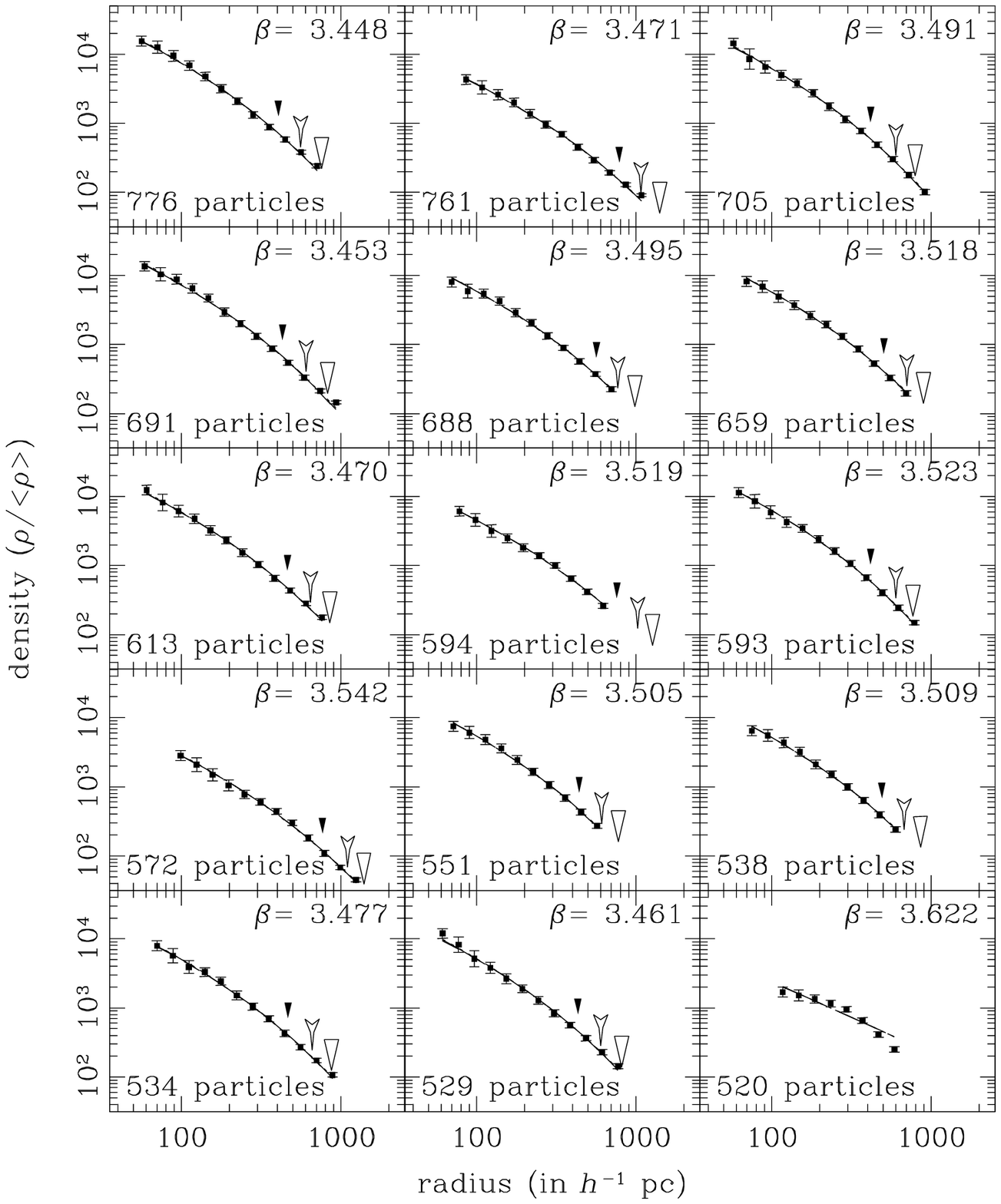,width=6in}
\caption{\label{profileb}Same as Fig.~\ref{profilea}.}
\end{figure}

We calculated the density profiles of the thirty most massive halos.  
The spherically averaged density as a function of radius
for these halos, with the 
potential minimum as the center, is plotted in Figs.~\ref{profilea} 
and \ref{profileb}.  The potential minimum was chosen over the 
center of mass because the potential minimum is more likely to be at 
the point of highest density.  

Possible analytic fits to dark matter halo profiles are those due
to Hernquist (1990) and Navarro, Frenk \& White (1997, henceforth NFW). 
Avila-Reese, \etal (1999) proposed the following 
general form for the fitting formulae: 
\begin{equation}
\rho(r) = \frac{\rho_0}{\frac{r}{r_s}\left(1+\frac{r}{r_s}\right)^{\beta-1}}
\end{equation}
where $\rho_0$ and $r_s$ are scalings for the density and radius, 
respectively, and $\beta$ is a power-law index parameter explained as follows.
When $\beta = 4$ this corresponds to a Hernquist profile, and when 
when $\beta = 3$ this corresponds to an NFW 
profile.  We can also allow $\beta$ to be a free 
parameter.  
The asymptotic behavior of the profiles is 
\begin{equation}
\rho \sim \left\{
\begin{array}{ll}
r^{-1},& r/r_s \ll 1 \\
r^{-\beta},& r/r_s \gg 1. \\
\end{array}
\right.
\end{equation}
Thus, $\beta$ parametrizes the profile shape at 
at radii much larger than the scale length.

The fits to the NFW and Hernquist profiles are indicated 
in Figs.~\ref{profilea} and \ref{profileb} by
dotted and short dashed lines, respectively.  A third model, allowing 
$\beta$ to be a free parameter in the fit, is indicated by 
a long dashed line.  This fitted $\beta$ is displayed at the bottom 
of each profile plot.  The scale radius $r_s$ for each of these three
profiles is indicated by filled, fat, and thin 
triangles, respectively.  

All three profiles are good fits to the 
data, and are indistiguishable.  This is because the power-law
slopes for all the halos are fairly close to $-2$.  Since each profile is a 
smoothly varying function, the power-law slope also varies smoothly 
from $-1$ to $-\beta$.  One can easily fit the part of the profile with 
power-law slope $-2$ to each halo profile.  
Note also that $\beta\sim3.5$ in all cases, intermediate to the 
NFW and Hernquist profile expectations.  
The close agreement of the fits can also be explained by the 
relatively large values of $r_s$ 
with respect to the sizes of the halos.  The halos are only a few 
scale radii in size, so we are well below the regime where 
the halo profiles behave as 
\(\rho \propto r^{-\beta}\).


\subsection{Spin parameter}
\label{spin}

\begin{figure}
\epsfig{file=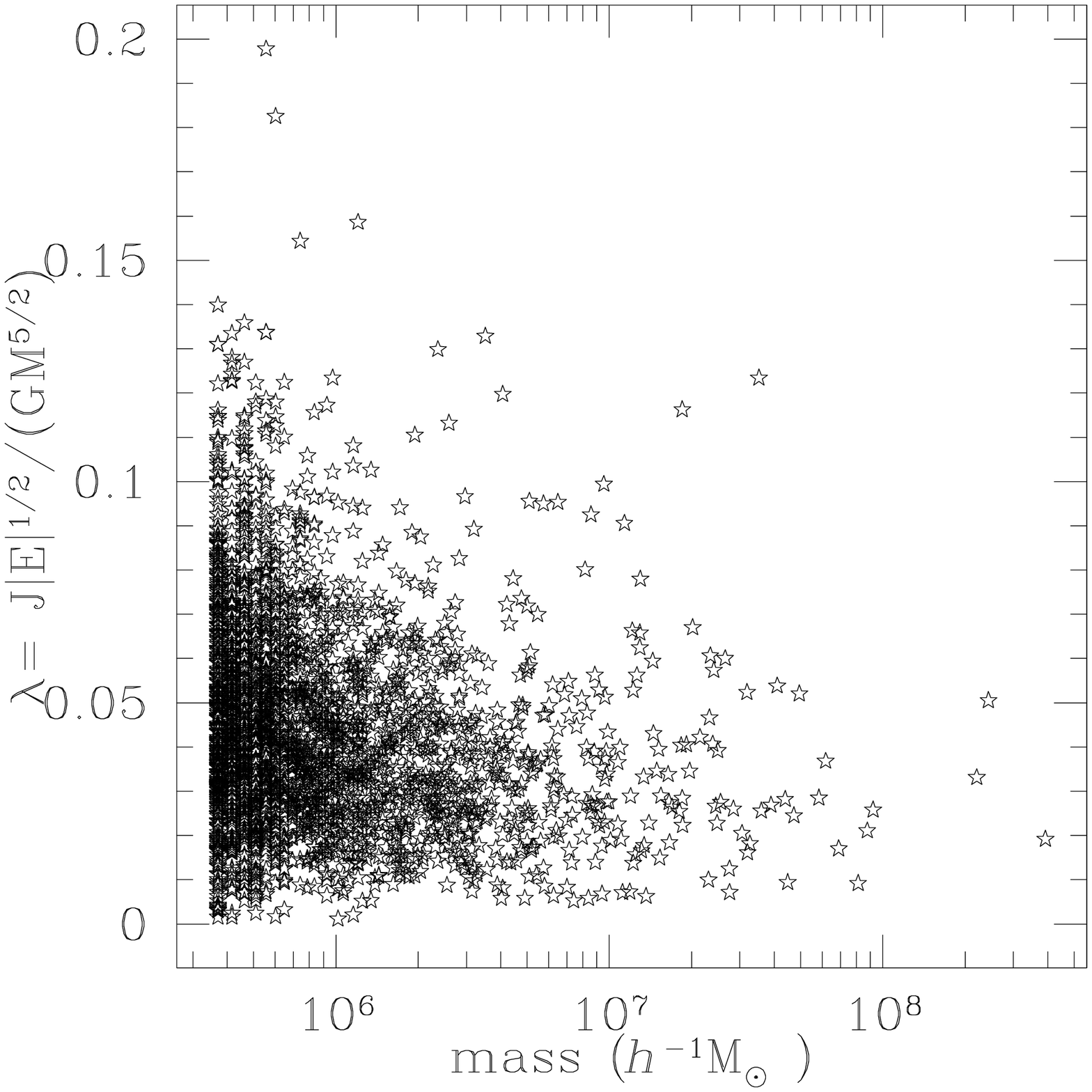,height=6in,width=6in}
\caption{\label{spinplot}Spin parameter versus mass.}
\end{figure}

The spin parameter of a bound object is defined as 
\begin{equation}
\lambda = \frac{J |E|^{1/2}}{G M^{5/2}}
\end{equation}
where $J$ is the angular momentum, $E=K+W$ is the binding energy, 
$G$ is the gravitational constant, and $M$ is the mass.  This is 
a dimensionless parameter that quantifies the amount of rotational support 
an object has.  A value of $\lambda \sim 1$ corresponds to nearly full
rotational support, 
and is typical of spiral galaxies, while $\lambda \sim 0.05$ is typical of 
elliptical galaxies, which are supported by velocity dispersion.  
Figure \ref{spinplot} is a plot of spin parameter versus mass for the 
halos in the simulation.  

\begin{figure}
\epsfig{file=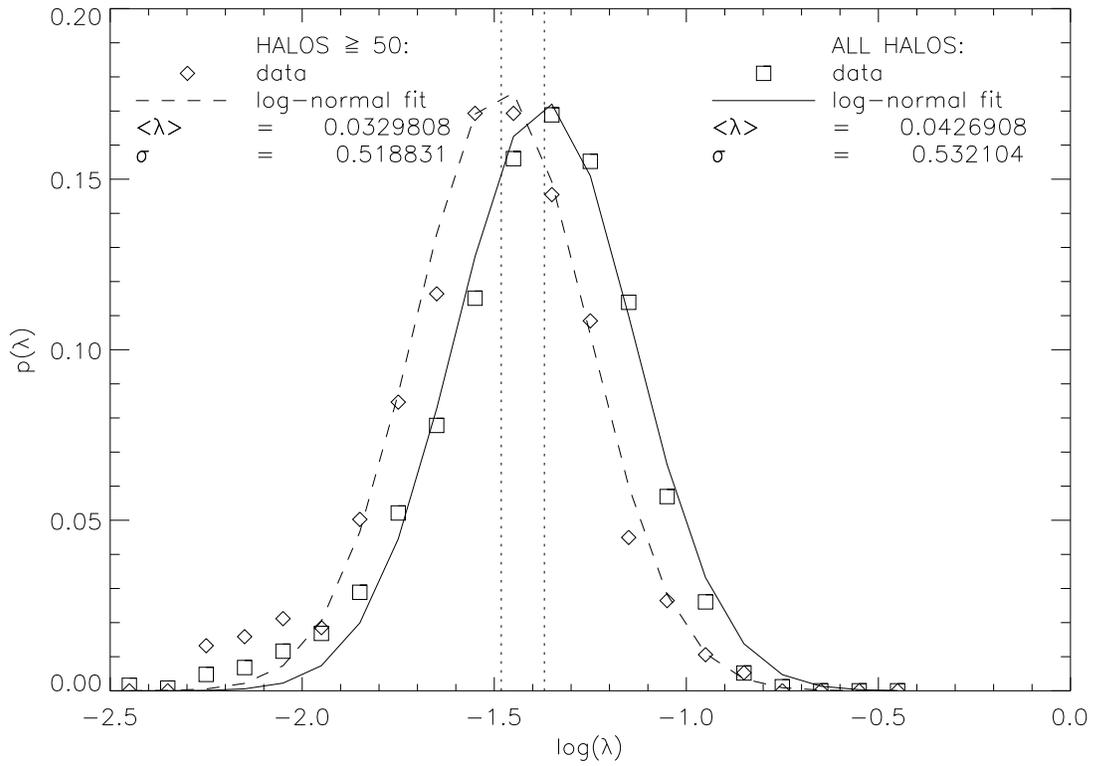,width=6in}
\caption{\label{spinlognormal}Distribution of spins.  The squares
and diamonds show the normalized distribution for all the halos and 
halos with at least 50 particles, respectively.  The solid line and 
the dashed line show the log-normal fit to the respective data sets.  
}
\end{figure}

Mo, Mao \& White (1998) found that the distribution of spin parameters 
of dark matter halos in galaxy formation simulations 
approximates a log-normal distribution, that is 
\begin{equation}
p(\lambda)\,d\lambda = \frac{1}{\sqrt{2\pi}\sigma}
	\exp\left[-\frac{\ln^2(\lambda/\bar{\lambda})}{2\sigma^2}\right]
	\,\frac{d\lambda}{\lambda}.  
\end{equation}
with fitting parameters 
of \(\bar{\lambda}=0.05\) and \(\sigma_{\lambda}=0.5\).  
The log-normal distribution is also a good fit for the data presented 
in this paper, as shown in Fig.~\ref{spinlognormal}.  The 
simulation data is displayed as squares, and the log-normal fit to it 
is the solid line.  
The fitting parameters are $\bar{\lambda}=0.043$ and 
$\sigma_\lambda=0.53$, which is consistent with the 
results of Mo, Mao \& White (1998).  

The smallest halos in the simulation contain only eight particles.  
For a halo with so few particles, the spin parameter is not a 
particularly meaningful quantity.  Since the most numerous halos in the 
simulation are the smallest ones, the distribution of $\lambda$ 
calculated above may not be particularly meaningful either.  
We recalculated the distribution of $\lambda$ for halos 
containing at least $50$ particles to determine the effect 
of excluding small halos.  This is plotted as diamonds in 
Fig.~\ref{spinlognormal}, and the log-normal fit to it is the dashed line. 
The fitting parameters for these 389 
halos is $\bar{\lambda}=0.033$ and $\sigma_\lambda=0.52$, indicating 
that the larger halos have systematically smaller spin parameters 
than smaller halos.

\subsection{Halo shapes}

A typical dark matter halo is not actually spherical, but is triaxial 
due to anisotropy in its velocity dispersion.  Thus, the shape of 
a halo can be quantified by finding its best fitting ellipsoid.  
The three axes of the best fitting ellipsoid 
will be referred to as the major, intermediate, and minor 
axes, in decreasing order of size.  

\subsubsection{Method of calculating ellipsoids}
\label{calcellip}

The best-fitting ellipsoid can be found by using the moment of inertia 
tensor, which is defined as 
\begin{eqnarray}
\label{MoITmatrix}
\mathbf{I} &=& 
\left[\begin{array}{ccc}
\sum\limits_i m_i (y_i^2+z_i^2) &	-\sum\limits_i m_i x_i y_i &	-\sum\limits_i m_i x_i z_i \\
-\sum\limits_i m_i y_i x_i &	\sum\limits_i m_i (x_i^2+z_i^2) &	-\sum\limits_i m_i y_i z_i \\
-\sum\limits_i m_i z_i x_i &	-\sum\limits_i m_i z_i y_i &	\sum\limits_i m_i (x_i^2+y_i^2)  \\
\end{array}\right] \nonumber \\
&=& \sum_i m_i \left(
\left[\begin{array}{ccc}
r_i^2 &	0 &	0 \\
0 &	r_i^2 &	0 \\
0 &	0 &	r_i^2 \\
\end{array}\right]
- 
\left[\begin{array}{ccc}
x_i^2  &	x_i y_i &	x_i z_i \\
y_i x_i &	y_i^2  &	y_i z_i \\
z_i x_i &	z_i y_i &	z_i^2   \\
\end{array}\right]
\right)
\nonumber \\
&=& \sum_i m_i (r_i^2\mathbf{1} - \mathbf{r}_i \mathbf{r}_i )
\end{eqnarray}
where 
\(r_i = \sqrt{x_i^2 + y_i^2 + z_i^2}\) is the distance of 
the $i$th particle to the 
center of the distribution of particles.  
The moment of inertia $I$ about an axis through the center of the distribution 
in the direction of the unit vector $\mathbf{n}$ 
is given by 
\begin{equation}
I = \mathbf{n}\cdot\mathbf{I}\cdot\mathbf{n}.
\end{equation}
Another property of the moment of inertia tensor is that the angular momentum 
$\mathbf{L}$ can be expressed as 
\begin{equation}
\mathbf{L} = \mathbf{I}\cdot\mathbf{\omega}.
\end{equation}
Thus, the eigenvectors correspond to the axes about which the 
angular velocity and angular momentum are aligned and the eigenvalues 
are the corresponding moments of inertia.

Consider an ellipsoid of constant density centered at the origin
with its axes along the $x$-, $y$-, and $z$-axes defined by 
\begin{equation}
\frac{x^2}{a^2} + \frac{y^2}{b^2} + \frac{z^2}{c^2} \leq 1.
\end{equation}
By symmetry, the 
moment of inertia tensor is already diagonalized, and the diagonal 
elements are the eigenvalues.  
It can be shown (through some rather tedious integration) that 
the diagonal elements are 
\begin{eqnarray}
I_1 &=& \frac M5 (b^2+c^2) \nonumber \\ 
I_2 &=& \frac M5 (a^2+c^2) \nonumber \\ 
I_3 &=& \frac M5 (a^2+b^2) \nonumber \\ 
\end{eqnarray}
where $M$ is the total mass of the ellipsoid.  
Using equation (\ref{MoITmatrix}), we now have
\begin{equation}
\frac M5 
\left[\begin{array}{ccc}
(b^2+c^2) & 0 & 0 \\
0 & (a^2+c^2) & 0 \\
0 & 0 & (a^2+b^2) \\
\end{array}\right]
=  \frac M5 (a^2+b^2+c^2) \mathbf{1} - \sum_i m_i \mathbf{r}_i \mathbf{r}_i
\end{equation}
which reduces to
\begin{equation}
\sum_i m_i \mathbf{r}_i \mathbf{r}_i = 
\frac M5 \left[\begin{array}{ccc}
a^2 & 0 & 0 \\
0 & b^2 & 0 \\
0 & 0 & c^2 \\
\end{array}\right].
\end{equation}
Thus, the axes of the best fitting ellipsoid can found by calculating the 
tensor 
\begin{equation}
M_{\alpha\beta} = \sum_i m_i x_{\alpha,i} x_{\beta,i}
\end{equation}
and finding its eigenvalues and eigenvectors.

The simplest way of determining the halo shape is to use the center of 
mass as the center of the distribution and to sum over all the particles 
in the halo to calculate the tensor $M_{\alpha\beta}$.  
However, some halos found by the {\tt skid} program include satellite halos 
or consist of two or more groups connected by a thin bridge.  The 
center of mass for these halos may not actually lie in the center of the 
halo.  In addition, ellipsoids calculated about the center of mass
have systematically higher 
axial ratios -- that is, they are more spheroidal in shape.  
This can be explained by the fact that {\tt skid} tends to calculate 
halos within a spherical volume, regardless of the intrinsic shape of the 
halo.  Thus, the calculated shape of the halo is more rounded out, so to 
speak.  

A better approximation is to use the potential 
minimum as the center of the distribution and to sum over the inner 
part of the halo.  We used an iterative method to fit an ellipsoid 
to half the mass of the halo.  The procedure begins by finding 
the sphere centered at the potential minimum that contains half the mass of 
the halo.  These particles are used to calculate an initial guess for the 
ellipsoid axes.  Keeping the orientation and axial ratios of the ellipsoid 
fixed, a new ellipsoid is calculated which contains half the mass of the 
halo.  The particles within this ellipsoid are used to calculate a new 
$M_{\alpha\beta}$ and a new guess for the ellipsoid is calculated.  
This procedure is 
repeated until the values of the axes converge.  
Dubinski and Carlberg (1991) employed a similar method to calculate halo 
shapes, but using particles within a fixed distance of the halo center 
rather than a fixed fraction of the mass.  
In order to leave out halos that are too small to have their shapes 
accurately calculated, we chose a lower cutoff to the halo size 
of 50 particles.  There were 389 halos of this size. 

\subsubsection{Ellipticities and triaxiality}

Using the iterative procedure described above, we calculated the 
magnitude and orientation of the principal axes of each halo.  
Henceforth, we shall refer to the lengths of the major, intermediate, 
and minor axes as $a$, $b$, and $c$, respectively.  
The ellipticities of the halos, 
\begin{eqnarray*}
\epsilon_1 &=& 1-b/a, \\
\epsilon_2 &=& 1-c/b,
\end{eqnarray*}
are useful measures of the shapes.  
A perfectly spherical halo would have $a=b=c$, hence
\(\epsilon_1=\epsilon_2=0.\) 
An oblate halo would have the larger two axes equal to each other 
($a=b$), hence \(\epsilon_1=0\) and \(\epsilon_2>0\).
A prolate halo would have the smaller two axes equal to each other
($b=c$), hence \(\epsilon_1>0\) and \(\epsilon_2=0\).
In general, however, a halo will be triaxial, meaning that 
there are no equalities between axis lengths.  

\begin{figure}
\epsfig{file=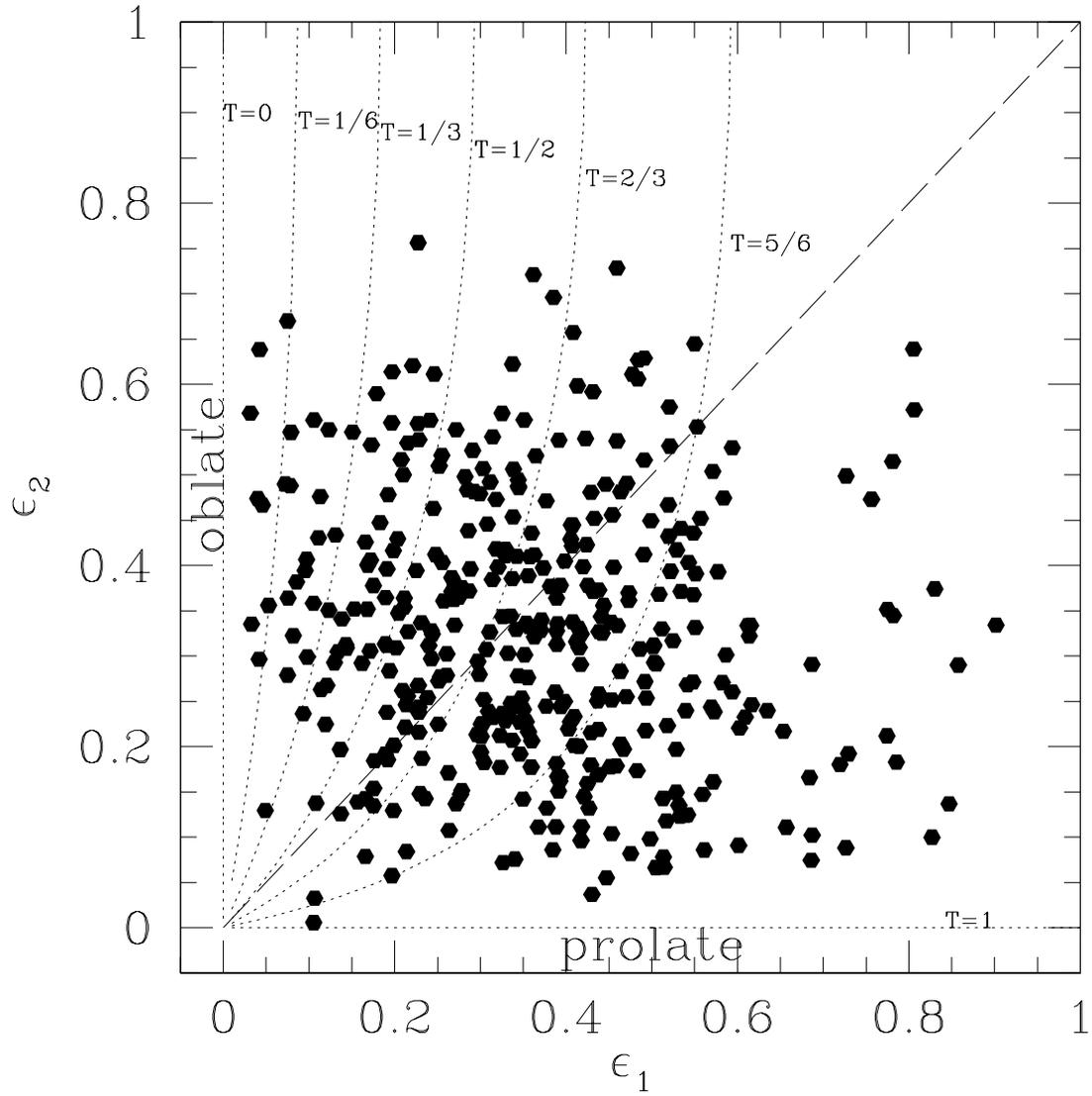,height=6in,width=6in}
\caption{\label{ellip}Ellipticities of halos of size $\geq50$ particles, 
defined as \(\epsilon_1=1-b/a\) and \(\epsilon_2=1-c/b\).
The dashed line shows the division between prolate and oblate halos, 
and the dotted lines show contours of constant triaxiality.
}
\end{figure}

Figure \ref{ellip} is a plot showing the ellipticities of the halos.  
The diagonal line shows the division between prolate and oblate halos -- 
prolate halos lie below the line, and oblate halos lie above it.  By 
this criterion, there are 212 prolate halos and 177 oblate halos.  
Warren \etal (1992) also found more prolate than 
oblate halos in their simulations of dark matter halos 
at zero redshift at galaxy size scales (the smallest mass they considered was
$3\times 10^9 M_{\sun}$).  They found this result to be independent of the 
initial power spectrum used, so it is reassuring that we 
obtain the same results for our simulation, although we use a different 
power spectrum and mass range.  

We can also compare the prolateness/oblateness of halos by use of 
the triaxiality parameter $T$, which is defined as 
\begin{equation}
T = \frac{a^2-b^2}{a^2-c^2}.
\end{equation}
A halo that is purely prolate has $b=c$, so $T=1$.  A halo that is 
purely oblate has $a=b$, so $T=0$.  The dotted lines in 
Fig.~\ref{ellip} represent contours of constant $T$.  

\begin{figure}
\epsfig{file=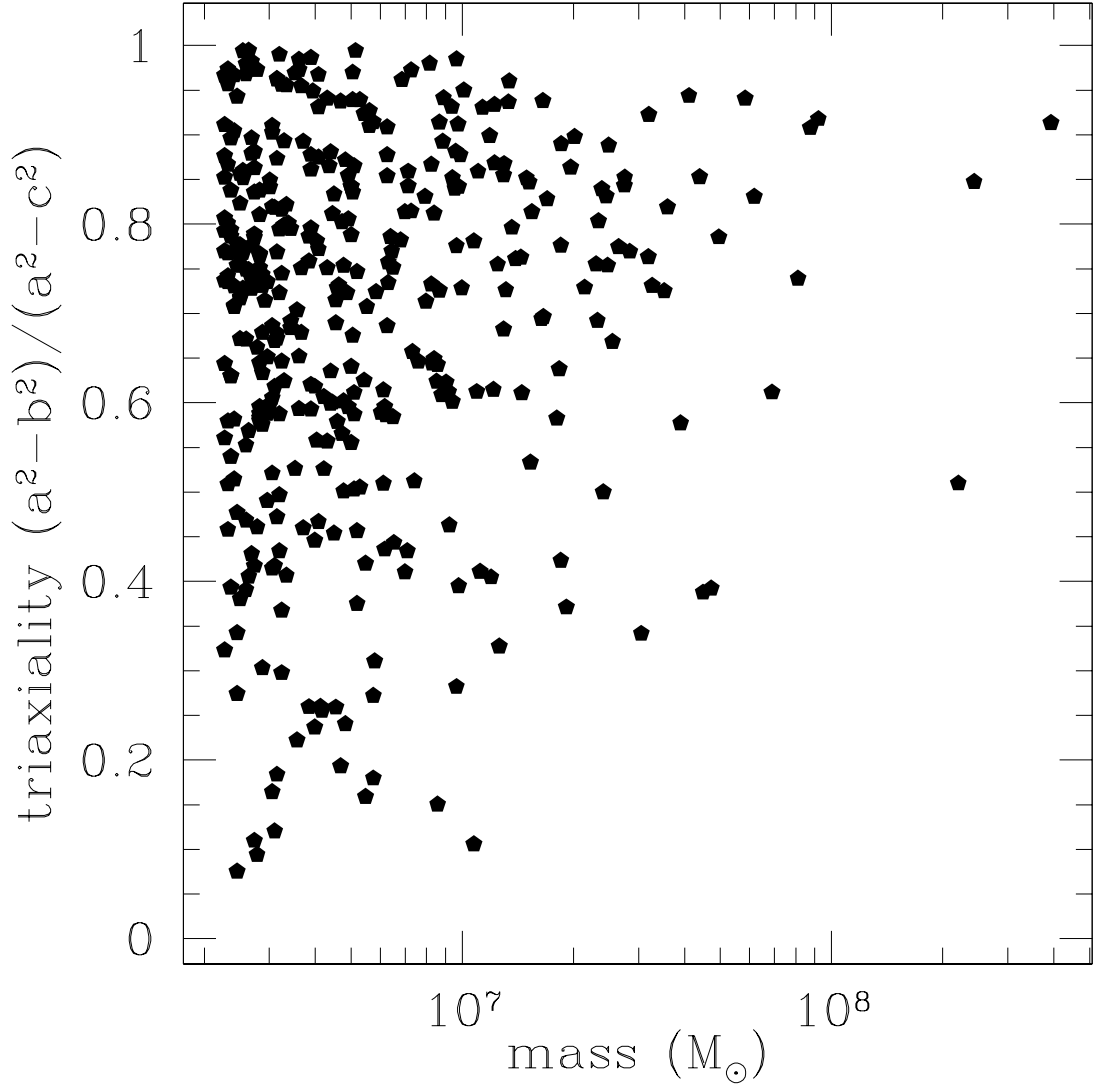,height=6in,width=6in}
\caption{\label{triaxiality}
Triaxiality of halos with $\geq50$ particles versus mass.
}
\end{figure}

\begin{figure}
\epsfig{file=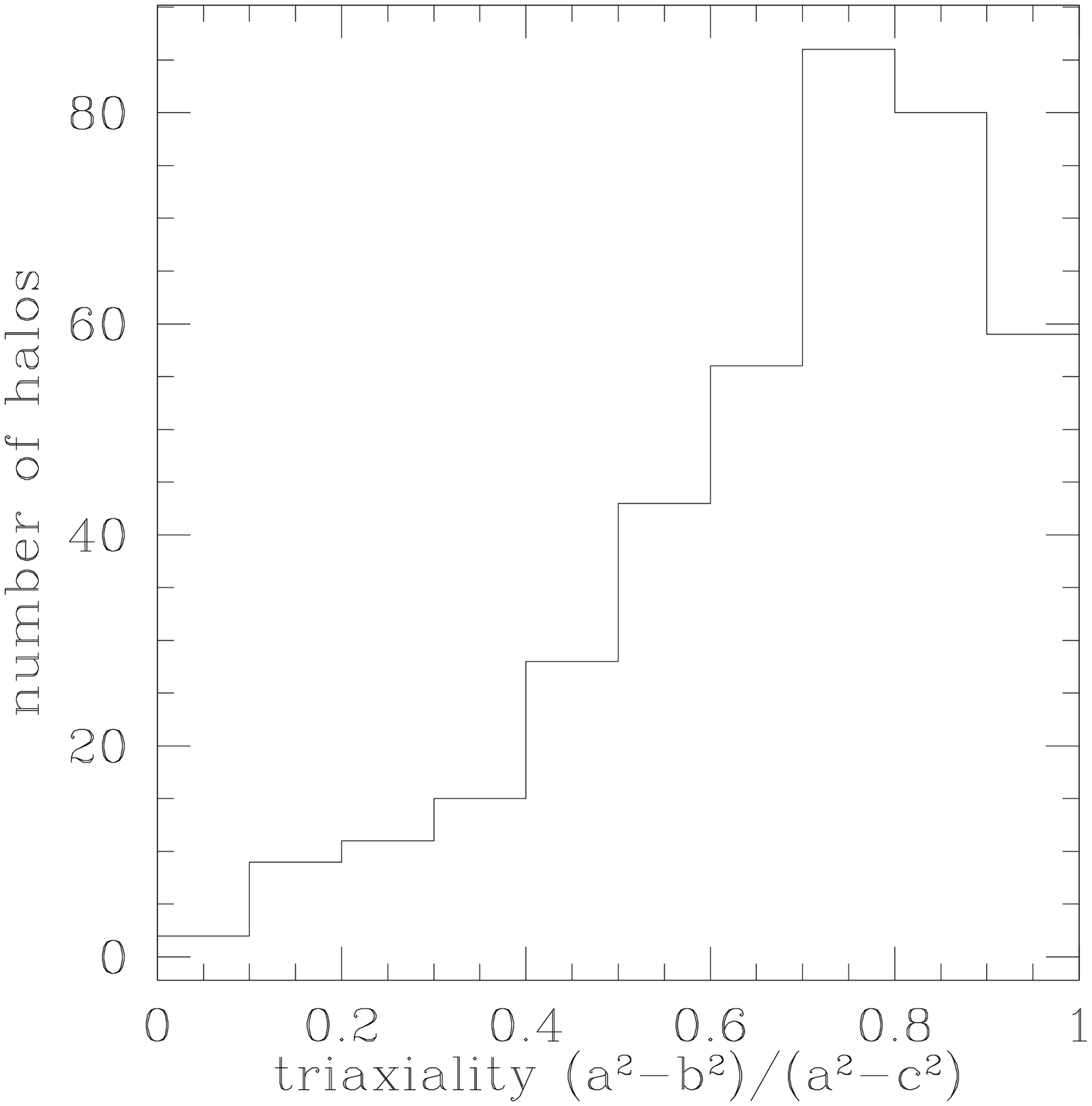,height=6in,width=6in}
\caption{\label{triax}Distribution of halo triaxiality parameter
$T$, defined as 
\(T = (a^2-b^2)/(a^2-c^2)\).  
}
\end{figure}

Figure \ref{triaxiality} 
is a plot of $T$ versus halo mass.  There does not appear to be any 
strong correlation between mass and triaxiality.  However, there is a 
tendency at all masses for the triaxiality to be close to $1$.  
Figure \ref{triax} shows the distribution of the triaxiality parameter 
for all halos with at least 50 particles.  The halos tend toward high 
values of $T$, another indication that prolate halos dominate.  
These two figures are qualitatively similar to Figs.~9a and 8 in 
Warren \etal (1992), indicating that our results are similar to theirs, 
despite the difference in mass scales and redshift.

\subsubsection{Angular momentum misalignment}

Now that we have calculated the axes of the best-fit ellipsoids of 
the halos, we can consider the orientation with respect to 
the angular momentum.  
Since the ellipsoids were fit using half the mass of the halo, 
we also recalculate the angular momenta using the 
positions and velocities of the same particles 
that were used to calculate the ellipsoids.
We find the cosine of the angles 
between the angular momentum and each of the principal axes 
by taking dot products.  

\begin{figure}
\epsfig{file=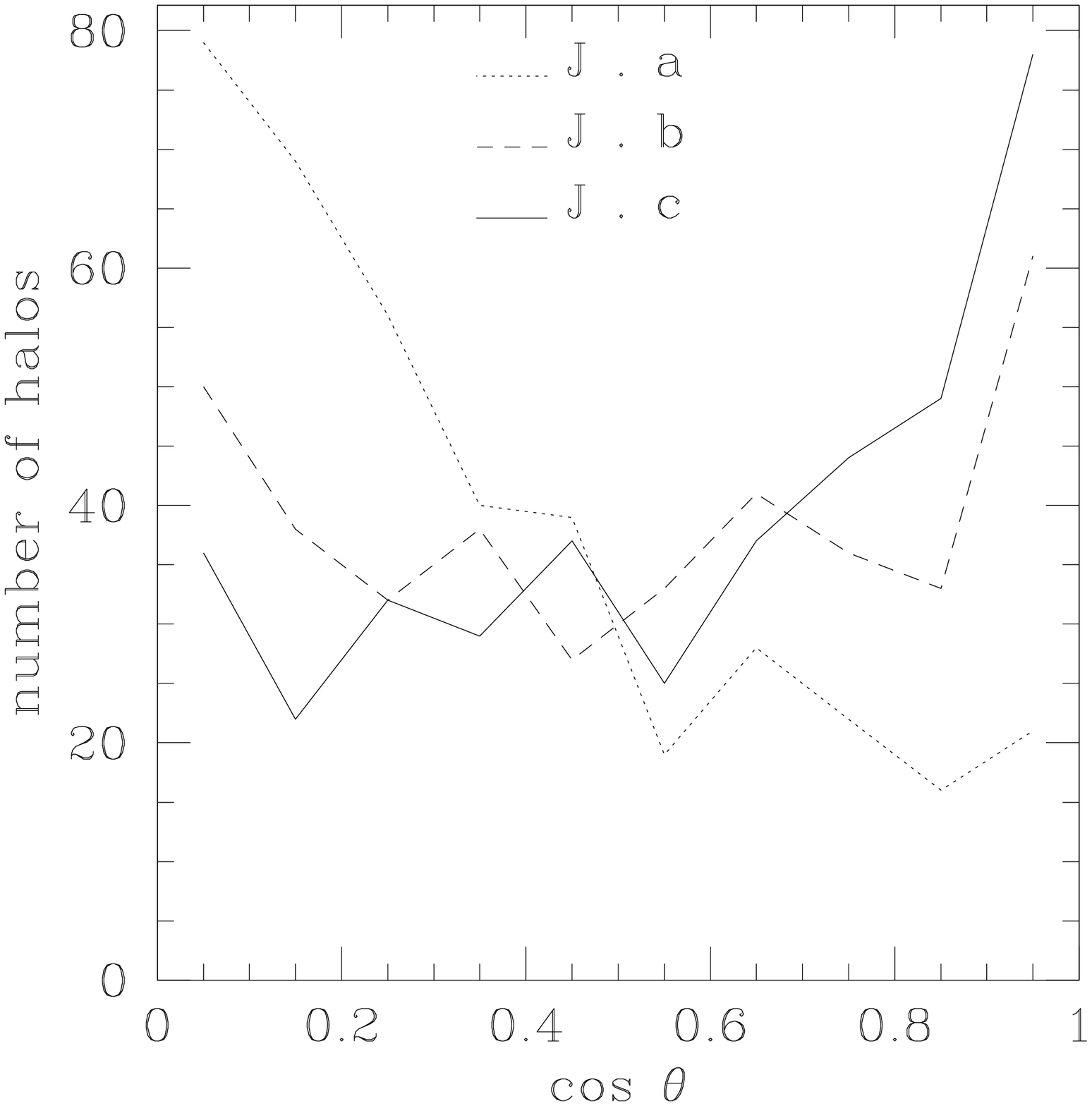,height=6in,width=6in}
\caption{\label{align}Distribution of angle cosines between angular 
momentum vector and ellipsoid axes for halos with $\geq50$ particles.  
The dotted (dashed, solid) line shows the cosine of the angle 
between the major (intermediate, minor) axis of the halo.
}
\end{figure}

Figure \ref{align} shows the distribution 
of these angle cosines. 
The dotted, dashed, and solid lines show the distribution of the 
angle cosines between the major, intermediate, and minor 
axes of the halo, respectively.  
As we can see in Fig.~\ref{align}, the angular momentum tends to 
be aligned with the minor axis of halos.
This is consistent with 
observations of elliptical galaxies which indicate that the 
angular momentum vector tends to align with the the minor axis 
(\cite{franx}).  

Kinematically, the angular momentum should align with either the 
major or minor axis because particles in a triaxial 
potential admit tube orbits only about these axes.  
In other words, particles may only circulate about either the major 
or the minor axis, causing the angular momentum of the system as a 
whole to align with these axes (\cite{BT}).  
However, most halos do not show alignment with the major axis.  In fact, 
many have their angular momentum perpendicular to their major axis.  

In addition, there are a number of halos whose angular momentum is aligned 
with the intermediate axis, even though this configuration is 
kinematically unstable.  
Warren \etal (1992) also observe this 
in their dark matter simulations, and suggest that this is caused by a 
long time scale for the realignment of orbits.

\section{Summary}

The major conclusion of that paper is that this N-body simulation of 
small scale structure formation at high redshift is similar to 
N-body simulations of large scale structure formation at at low redshift.
The details can be summarized as follows:
\begin{enumerate}
\item The mass function of the halos can be well approximated by the 
Press-Schechter formalism.  
\item The profile of the halos can be modeled equally well by either the 
NFW and Hernquist profiles.
\item The shapes of the halos are generally triaxial, with a tendency 
toward prolateness.
\item The average spin parameter of the halos is about $0.04$. 
\item The angular momentum tends to align with the minor axis most often, 
and favors alignment with the intermediate axis over the major axis.  
\end{enumerate}

In fact, it would be surprising if the results were 
completely different from other N-body simulations, 
since they all model the gravitational 
collapse of objects from some primordial power spectrum of density 
fluctuations.  
The main difference is in the shape of the power spectrum, which 
changes as we go to smaller scales.  This produces a corresponding 
change in the mass spectrum of objects that are collapsing.  

The Press-Schechter formalism predicts how the mass spectrum of 
collapsed objects depends on the initial power spectrum.  
We find that the mass function matches the predictions of Press-Schechter 
remarkably well, even at the low mass end.  This is consistent with 
previous work with N-body simulations at low redshift and large scales
which have also shown good agreement with the Press-Schechter formalism 
(\cite{EFWD}; \cite{kauffmanwhite}; \cite{laceycole}; \cite{VLS}).

The remaining halo properties described in this paper, including 
density profile, halo shape, and spin parameter, all are consistent with 
previous work with N-body simulations regardless of the power spectrum 
used.  Some authors use scale invariant power spectra, 
but using a range of spectral indices: 
choices of $-2$, $-1$, and $0$ are typical
(\cite{colelacey}; \cite{warren}).  Some use a CDM spectrum 
or some variant (\cite{MMW}).  Note that at the high redshifts 
studied in this paper, 
the dynamical evolution of a $\Lambda$CDM universe is 
the same as that of a pure CDM universe, since the matter density 
varies as $(1+z)^3$ while the vacuum energy density remains constant.  
However, the spectral index approaches $\sim -3$ on small scales, 
so we effectively choose a different power spectrum by considering 
small scales.  Although the initial power spectrum used 
in this paper is different from previous work, the overall results 
on halo properties remains the same.  

The shapes of the density profiles of the halos are consistent with 
both NFW and Hernquist profiles, which are based on studies of collapsed 
halos in cold dark matter simulations.
Indeed, when the shape of the outer 
profile $\beta$ is allowed to be a free parameter, we find that 
\(3 < \beta < 4\), intermediate to the NFW and Hernquist profiles.  

The results for the spin parameter and shapes of halos are also 
consistent with large-scale structure simulations.  
The median and distribution of the spin parameter are both 
similar to the those found for large-scale simulations, 
i.e.~\(\bar{\lambda}=0.043\), with a log-normal distribution
(\cite{MMW}; \cite{colelacey}).
The predominance of prolate halo shapes also agrees with these other 
simulations (\cite{warren}; \cite{colelacey}), as well as the alignment 
of the angular momentum with the minor axis (\cite{weil}; \cite{warren}).  
We also find that the angular momentum favors alignment with the 
intermediate axes over the major axis, as Warren \etal (1992) do. 

We can infer from this simulation that dark matter halos on small 
scales at high redshift behave very similarly to halos on large 
scales at low redshift.  This can help us understand high-redshift 
star formation by providing information on the dark matter 
environments in which the first stars formed.  Further study 
using simulations such as this may shed light on the IMF of the 
first stars, and their subsequent fate.
By adding additional 
physics to the simulation, such as gas physics and radiative transfer, 
we can study the nature of the first objects 
that formed in the universe.
\newline\

This work was supported under an NSF Graduate Fellowship, by NASA
Astrophysical Theory Grant NAG5-8727, and by NSF grants ACI96-19019
and AST-9803137.

\end{document}